\documentstyle[epsfig,12pt]{article}
\begin{document}
\begin{large}
\centerline{ A study of pentaquark $\Theta$ state in the  chiral SU(3)
quark model
\footnote{Project supported by the National Natural Science
Foundation of China }}
\end{large}
\par
\vspace{0.5cm} \centerline{F. Huang, Z.Y. Zhang, Y.W. Yu}
\begin{small}
\centerline{Institute of High Energy Physics,  Beijing 100039,
P.R.China}
\end{small}
\bigskip
\centerline{B.S.Zou}
\begin{small}
\centerline{CCAST (World Laboratory), P.O.~Box 8730, Beijing
100080;} \centerline{Institute of High Energy Physics, Beijing
100039, P.R.China}
\end{small}
\vskip 0.5in

\begin{abstract}
The structure of the pentaquark state $uudd$-$\bar{s}$ is studied
in the chiral $SU(3)$ quark model as well as in the extended
chiral $SU(3)$ quark model, in which the vector meson exchanges
are included. Four configurations of $J^{\pi}=\frac{1}{2}^-$ and
four of $J^{\pi}=\frac{1}{2}^+$ are considered. The results show
that the isospin $T=0$ state is always the lowest one for both
$J^{\pi}=\frac{1}{2}^-$ and $J^{\pi}=\frac{1}{2}^+$ cases in
various models. But the theoretical value of the lowest one is
still about $200 - 300MeV$ higher than the experimental mass of
$\Theta$. It seems that a dynamical calculation should be done for
the further study.

\end{abstract}

\vspace{1.0cm}



\vspace{1.5cm}
\section{Introduction}

Recently, LEPS Collaboration at SPring 8 \cite{s1}, DIANA Collaboration
at ITEP \cite{s2}, CLAS Collaboration at Jefferson Lab \cite{s3}
and SAPHIR Collaboration at ELSA \cite{s4} report that they observed
a new resonance $\Theta$, with strangeness quantum number $ S=+1 $.
The mass of this $\Theta$ particle is around $M_{\Theta}=1540 MeV$ and
the upper limit of the width is about $\Gamma_{\Theta}< 25 MeV$.
Since it has strangeness quantum number $ S=+1 $, it must be a
5-quark system. The interesting problem is  whether it is a
strange meson-baryon molecule like state or a pentaquark state.
If it is really a pentaquark state, it will be the first
multi-quark state people found. There are already many theoretical
works to try to explain its properties with various quark models
\cite{s5,s6,s7} or other approaches \cite{s8} .
A re-analysis \cite{s9} of older experimental data on the
$K^+$-nucleon elastic scattering process put a more stringent constraint
on the width to be $\Gamma_{\Theta}< 1 MeV$.  Since the mass
of $\Theta$, $M_{\Theta}$, is larger than the sum of nucleon mass and kaon mass,
$ M_N + M_K $, it is not easy to understand why its width is so narrow, unless
it has very special quantum numbers. As to the mass of $\Theta$,
although it is predicted by the original chiral soliton model \cite{s10}
quite well, there is no concrete calculation from quark model
available yet.

In this work, we calculate the energies of the pentaquark states
in chiral quark model.  Four configurations of
$J^{\pi}=\frac{1}{2}^-$ and four of $J^{\pi}=\frac{1}{2}^+$ are
considered. Some qualitative information is obtained: (1) The
isoscalar state, $ T=0 $, is always the lowest one for both cases
of $J^{\pi}=\frac{1}{2}^-$ and $J^{\pi}=\frac{1}{2}^+$. (2) The
calculated results of the extended chiral $SU(3)$ quark model are
quite similar to those of the chiral $SU(3)$ quark model, when the
parameters are taken as what we used in the $N-N$ scattering
calculation \cite{s11,s12}. (3) When the size parameter is
adjusted to be $ 0.6 fm $, the energy of the lowest state
$~([4]_{orb}[31]_{ts=01}^{\sigma
f}\bar{s},LST=0\frac{1}{2}0,~J^{\pi}=\frac{1}{2}^-)~$, is $1670
MeV$, still about $130 MeV$ higher than the experimental value of
the $\Theta$ mass. A dynamical calculation will be done for
getting quantitative information of the $\Theta$ particle's
structure.

\section{Theoretical framework}
For a $4q-\bar{q}$ color singlet system, the $4q$ wave function
includes three parts: orbital, flavor-spin $SU(3)\times SU(2)$ and
color $SU(3)$ part. In $\Theta$ particle case, its strangeness is
$+1$, $4q$ part only includes u and d quarks, and the anti-quark
is $\bar{s}$. Four configurations for $J^{\pi}=\frac{1}{2}^-$ are
considered, they are: $~(~[4]_{orb}~[31]_{ts=01}^{\sigma
f}~\bar{s},~LST=0\frac{1}{2}0,~J^{\pi}=\frac{1}{2}^-~)~$,
$~(~[4]_{orb}~[31]_{ts=10}^{\sigma
f}~\bar{s},~LST=0\frac{1}{2}1,~J^{\pi}=\frac{1}{2}^-~)~$,
$~(~[4]_{orb}~[31]_{ts=11}^{\sigma
f}~\bar{s},~LST=0\frac{1}{2}1,~J^{\pi}=\frac{1}{2}^-~)~$ and
$~(~[4]_{orb}~[31]_{ts=21}^{\sigma
f}~\bar{s},~LST=0\frac{1}{2}2,~J^{\pi}=\frac{1}{2}^-~)~$. We also
considered 4 configurations for $J^{\pi}=\frac{1}{2}^+$:
$~(~[31]_{orb}~[4]_{ts=00}^{\sigma
f}~\bar{s},~LST=1\frac{1}{2}0,~J^{\pi}=\frac{1}{2}^+~)~$,
$~(~[31]_{orb}~[4]_{ts=11}^{\sigma
f}~\bar{s},~LST=1\frac{1}{2}1,~J^{\pi}=\frac{1}{2}^+~)~$,
$~(~[31]_{orb}~[4]_{ts=11}^{\sigma
f}~\bar{s},~LST=1\frac{3}{2}1,~J^{\pi}=\frac{1}{2}^+~)~$ and
$~(~[31]_{orb}~[4]_{ts=22}^{\sigma
f}~\bar{s},~LST=1\frac{3}{2}2,~J^{\pi}=\frac{1}{2}^+~)~$. The
color part of them is $[211]^c$, i.e.  $(\lambda \mu)_c =(10)$,
combining $(01)$ of $\bar{s}$, the total quantum number in color
space is singlet. For $J^{\pi}=\frac{1}{2}^-$ states, color
$[211]^c$ with spin-flavor $[31]^{\sigma f}$ constructs the total
anti-symmetric structure of the $4q$ part, and for
$J^{\pi}=\frac{1}{2}^+$ states, $[31]_{orb}$ replaces
$[31]^{\sigma f}$ to make the anti-symmetrization.

In the chiral $SU(3)$ quark model the Hamiltonian of the system can be written as
\begin{eqnarray}
& H & =\sum\limits_{i}T_i-T_G+\sum\limits_{i<j=1-4}V_{ij}+\sum\limits_{i=1-4}V_{i5},
\end{eqnarray}
where ~$\sum\limits_{i}T_i-T_{G}$ is the kinetic energy of the
system, $V_{ij}, i,j=1-4$ and $V_{i5},i=1-4$ represent the
interactions between quark-quark $(q-q)$ and quark-anti-quark $(q-\bar{q})$
respectively.
\begin{eqnarray}
& V_{ij} & =V_{ij}^{conf}+V_{ij}^{OGE}+V_{ij}^{ch},
\end{eqnarray}
$V_{ij}^{conf}$ is the confinement potential taken as the
quadratic form,
\begin{eqnarray}
& V_{ij}^{conf} &
=-a_{ij}^{c}(\lambda_{i}^{c}\cdot\lambda_{j}^{c})r_{ij}^2
-a_{ij}^{c0}(\lambda_{i}^{c}\cdot\lambda_{j}^{c}),
\end{eqnarray}
and $V_{ij}^{OGE}$ is the one gluon exchange (OGE) interaction,
\begin{eqnarray}
& V^{OGE}_{ij} & = \frac{1}{4} g_{i}g_{j} (\lambda^{c}_{i} \cdot
       \lambda^{c}_{j})
       \{ \frac{1}{r_{ij}} - \frac{\pi}{2} \delta(\vec{r}_{ij})
       ( \frac{1}{m^{2}_{qi}} + \frac{1}{m^{2}_{qj}} \nonumber \\
   & & + \frac{4}{3} \frac{1}{m_{qi}m_{qj}} (\vec{\sigma_{i}}
        \cdot \vec{\sigma_{j}}) )
\} + V_{tensor}^{OGE}
  + V_{\vec{\ell}\cdot\vec{s}}^{OGE}
\end{eqnarray}
$V_{ij}^{ch}$ represents the interactions from chiral field
couplings. In the  chiral $SU(3)$ quark model
$V_{ij}^{ch}$ includes scalar meson exchange $V_{ij}^{s}$ ,
pseudo-scalar meson exchange $V_{ij}^{ps}$, and in the extended
chiral $SU(3)$ quark model, vector meson
exchange $V_{ij}^{v}$ potentials are also included,
\begin{eqnarray}
& V_{ij}^{ch} & = \sum^{8}_{a=0} V_{s_a} (\vec{r}_{ij}) +
\sum^{8}_{a=0}
V_{ps_a} (\vec{r}_{ij})+ \sum^{8}_{a=0}V_{v_a} (\vec{r}_{ij})~.
\end{eqnarray}
Their expressions can be found in Refs.\cite{s11,s12}. The
interaction between $q$ and $\bar{q}$ includes two parts: direct
interaction and annihilation part,
\begin{eqnarray}
V_{i5} = V_{q\bar{q}}^{dir}+ V_{q\bar{q}}^{ann},
\end{eqnarray}
\begin{eqnarray}
V_{q\bar{q}}^{dir} = V_{q\bar{q}}^{conf}+ V_{q\bar{q}}^{OGE}+ V_{q\bar{q}}^{ch},
\end{eqnarray}
with
\begin{eqnarray}
V_{q\bar{q}}^{ch}(\vec{r}) =\sum_{i}(-1)^{G_i}V_{qq}^{ch,i}(\vec{r}) .
\end{eqnarray}
Here $(-1)^{G_i}$ describes the G parity of the ith meson.
For the $\Theta$ particle case, $q\bar{q}$ can only annihilate
into $K$ and $K^*$ mesons, thus $V_{i5}^{ann}$ can be expressed as:
\begin{eqnarray}
V_{i5}^{ann} = V_{ann}^{K}+ V_{ann}^{K^*},
\end{eqnarray}
with
\begin{eqnarray}
& V_{ann}^{K} = & \tilde{g}_{ch}^2 ~ \frac{1}{(\tilde{m}+\tilde{m}_{s})^2-m_K^2} ~
(\frac{1-\vec{\sigma}_q \cdot \vec{\sigma}_{\bar{q}}}{2})_{spin}
~~(\frac{2 + 3\lambda_q \cdot \lambda^*_{\bar{q}}}{6})_{color} \nonumber \\
& &(\frac{19}{9} + \frac{1}{6} \lambda_q \cdot \lambda^*_{\bar{q}})_{flavor}
~~\delta(\vec{r}_q -\vec{r}_{\bar{q}}),
\end{eqnarray}
and
\begin{eqnarray}
& V_{ann}^{K^*} = &\tilde{g} _{chv}^2 ~ \frac{1}{(\tilde{m}+\tilde{m}_{s})^2-m_{K^*}^2}~
(\frac{3+\vec{\sigma}_q \cdot \vec{\sigma}_{\bar{q}}}{2})_{spin}
~~(\frac{2 + 3\lambda_q \cdot \lambda^*_{\bar{q}}}{6})_{color} \nonumber \\
& &(\frac{19}{9} + \frac{1}{6} \lambda_q \cdot \lambda^*_{\bar{q}})_{flavor}
~~\delta(\vec{r}_q -\vec{r}_{\bar{q}}).
\end{eqnarray}
Where $\tilde{g}_{ch}$ and $\tilde{g}_{chv}$ are the coupling constants of
pseudo-scalar-scalar chiral field and vector chiral field in the
annihilation case respectively. $\tilde{m}$ represents the effective quark mass.
Actually, $\tilde{m}$ is quark momentum dependent, here we treat
it as an effective mass.

Using these two models, we did an adiabatic approximation
calculation to study the energies of the $(uudd$-$\bar{s})$
system.
\section{Results and discussions}
First, we carry on the calculation by taking the parameters which
can reasonably reproduce the experimental data of $N-N$ and $Y-N$
scattering \cite{s11,s12}. In the chiral $SU(3)$ quark model,
besides pseudo-scalar and scalar fields coupling, $OGE$
interaction is still there to offer part of the short range
repulsion, as well as in the extended chiral $SU(3)$ quark model,
the $OGE$ interaction is almost replaced by the vector meson
exchanges. About the annihilation interaction between
$u(d)-\bar{s}$, it is a complicated problem, in Eqs. (10) and
(11), the quark effective masses $\tilde{m}$ and $\tilde{m}_s$ ,
as well as the annihilation coupling constants $\tilde{g}_{ch}$
and $\tilde{g}_{chv}$ are subject to significant uncertainties. In
our calculation, we treat $(\tilde{m}+\tilde{m}_s)$,
$\tilde{g}_{ch}$ and $\tilde{g}_{chv}$ as parameters, and adjust
them to fit the masses of $K$ and $K^*$ mesons, named case I.  In
case II, we omitted the annihilation interaction in the
calculation to see its effects. All results of 4 configurations of
$J^{\pi}=\frac{1}{2}^-$ and 4 of $J^{\pi}=\frac{1}{2}^+$ are
listed in Table 1.

\vspace{0.5cm}

>From Table 1, one can see that: (1) The isoscalar state $(T=0)$ is
always the lowest state both in $ J^{\pi}=\frac{1}{2}^-$ and
in $ J^{\pi}=\frac{1}{2}^+$ cases, and
$~(~[4]_{orb}~[31]_{ts=01}^{\sigma f}~\bar{s},~LST=0\frac{1}{2}0,~J^{\pi}=\frac{1}{2}^-~)~$
is always the lowest one in different models.
(2) The results of the chiral $SU(3)$ quark model and the extended chiral $SU(3)$
quark model are quite similar, although the short range interactions
of these two models are different, one is from $OGE$ and the other
is from vector meson exchanges. (3) The annihilation interactions
offer attraction to the states of $ J^{\pi}=\frac{1}{2}^-$  and
repulsion to $ J^{\pi}=\frac{1}{2}^+$ states. (4) When the
annihilation interaction is considered,  the energy of the lowest state,
$~(~[4]_{orb}~[31]_{ts=01}^{\sigma f}~\bar{s},~LST=0\frac{1}{2}0,~J^{\pi}=\frac{1}{2}^-~)~$,
is about $250-300 MeV$ higher than the experimental value
of the $\Theta$ mass.

\begin{table}
\caption{Energies of pentaquark states in different chiral quark
model }
\vspace{1.5cm}
\begin{center}
\begin{tabular}{|c|c|c|c|}
\hline
                &   Chiral $SU(3)$ Quark Model & Ex. Chiral $SU(3)$ Quark Model \\
configuration   &  $b_u$=0.50 fm               & $b_u$=0.45 fm                  \\ \hline
$ J^{\pi}=\frac{1}{2}^-$                   & I~~~~~~~~~~~~II         & I~~~~~~~~~~~~II           \\
                                           &   $(MeV)$               &   $(MeV)$     \\ \hline
$[4]_{orb}[31]^{\sigma f}_{ts=01}~\bar{s}$  & 1801 ~~~~~~1957         & 1843 ~~~~~~2091           \\
$[4]_{orb}[31]^{\sigma f}_{ts=10}~\bar{s}$  & 2049 ~~~~~~2128         & 2089 ~~~~~~2170           \\
$[4]_{orb}[31]^{\sigma f}_{ts=11}~\bar{s}$  & 2117 ~~~~~~2190         & 2115 ~~~~~~2193           \\
$[4]_{orb}[31]^{\sigma f}_{ts=21}~\bar{s}$  & 2323 ~~~~~~2369         & 2314 ~~~~~~2334           \\
\hline
$ J^{\pi}=\frac{1}{2}^+$                   & I~~~~~~~~~~~~II         & I~~~~~~~~~~~~II           \\
                                           &   $(MeV)$               &   $(MeV)$     \\ \hline
$[31]_{orb}[4]^{\sigma f}_{ts=00}~\bar{s}$  & 2271 ~~~~~~2185         & 2270 ~~~~~~2253           \\
$[31]_{orb}[4]^{\sigma f}_{ts=11}~\bar{s}$  & 2308 ~~~~~~2235         & 2296 ~~~~~~2310           \\
~~~~~~~~~~~~~~~~~$(S=\frac{1}{2})$         &                         &                           \\
$[31]_{orb}[4]^{\sigma f}_{ts=11}~\bar{s}$  & 2362 ~~~~~~2282         & 2367 ~~~~~~2337           \\
~~~~~~~~~~~~~~~~~$(S=\frac{3}{2})$         &                         &                           \\
$[31]_{orb}[4]^{\sigma f}_{ts=22}~\bar{s}$  & 2426 ~~~~~~2367         & 2412 ~~~~~~2435           \\
\hline
\end{tabular}
\end{center}
\end{table}
\vspace{0.5cm}

We tried to adjust the size parameter $b_u$ to be larger to see
the influence. As an example, the results of $b_u=0.6 fm$
in the chiral $SU(3)$ quark model are given in Table 2.
In this case, the energies of all states become smaller,
caused by the kinetic energy of the system is reduced
for larger $b_u$. When the annihilation interaction is
included (case I), the energy of the lowest state,
$~(~[4]_{orb}~[31]_{ts=01}^{\sigma f}~\bar{s},~LST=0\frac{1}{2}0,~J^{\pi}=\frac{1}{2}^-~)~$,
is $1670 MeV$, about $130 MeV$ higher than the $\Theta$' mass.

\vspace{0.5cm}

In our results, the states of $ J^{\pi}=\frac{1}{2}^-$ are always
lower than those of $ J^{\pi}=\frac{1}{2}^+$, even in the extended
chiral $SU(3)$ quark model, in which the $OGE$ interaction is
almost totally replaced by vector meson exchanges. According to
Stancu and Riska's argument \cite{s6}, the state of $T=0,
J^{\pi}=\frac{1}{2}^+$ can be lower than the state of $T=0,
J^{\pi}=\frac{1}{2}^-$, because the spin-flavor dependent
interactions from Goldstone-Boson exchange potential offer more
attractions to the state of $T=0, J^{\pi}=\frac{1}{2}^+$. In our
calculation, it is true that $\pi $ and $\rho$ meson exchanges do
contribute very strong attractions to the state of $T=0,
J^{\pi}=\frac{1}{2}^+$, but when the interactions between $u(d)$
and $\bar{s}$ are included, especially the annihilation terms are
considered, the state of $T=0, J^{\pi}=\frac{1}{2}^-$ gets more
attractions. This is because that among $4$ pairs $u(d)-\bar{s}$
interactions, the state of $T=0, J^{\pi}=\frac{1}{2}^-$ has $1$
pair $u-\bar{s}$ of $(0s)^2$ with spin $s=0$ and color singlet
$(00)_c$ (i.e. $K$ meson's quantum numbers) and $\frac{1}{3}$ pair
of $(0s)^2$~$s=1$ ~$(00)_c$, the other part is color octet, but
the state of $T=0, J^{\pi}=\frac{1}{2}^+$ only has $\frac{1}{12}$
pair of $(0s)^2$~$s=0$~$(00)_c$, $\frac{1}{4}$ pair of
$(0s0p)$~$s=0$~$(00)_c$, $\frac{1}{4}$ pair of
$(0s)^2$~$s=1$~$(00)_c$, $\frac{3}{4}$ pair of
$(0s0p)$~$s=1$~$(00)_c$ and the other part is color octet. If we
take the annihilation interaction to fit the masses of $K$ and
$K^*$, the state of $T=0, J^{\pi}=\frac{1}{2}^-$ must be the
lowest. In Table 2, case II is of the annihilation interactions
omitted, and the results of without any interactions between $4q$
and $\bar{s}$ are given in column III. One sees from Table 2 that
when the interactions between $4q$ and $\bar{s}$ are all omitted,
the state of $T=0, J^{\pi}=\frac{1}{2}^+$ can be the lowest one,
but its energy $(1997 MeV)$ is much higher than the $\Theta$'s
mass.

\begin{table}
\caption{Energies of pentaquark states in chiral $SU(3)$ quark model with $b_u=0.6 fm$}
\begin{center}
\begin{tabular}{|c|c|c|}
\hline
                                                 & Chiral $SU(3)$ Quark Model  \\
configuration                                    & $(b_u=0.60 fm )$            \\ \hline
$ J^{\pi}=\frac{1}{2}^-$                         & ~~~~I~~~~~~~~~~~~~~II~~~~~~~~~~~~~~III~~~~ \\
                                                 &       $(MeV)$               \\ \hline
$[4]_{orb}[31]^{\sigma f}_{ts=01}~\bar{s}$        & ~~~~1672 ~~~~~~~~~1867 ~~~~~~~~~~~2027~~~~ \\
$[4]_{orb}[31]^{\sigma f}_{ts=10}~\bar{s}$        & ~~~~1940 ~~~~~~~~~1990 ~~~~~~~~~~~2039~~~~ \\
$[4]_{orb}[31]^{\sigma f}_{ts=11}~\bar{s}$        & ~~~~1983 ~~~~~~~~~2026 ~~~~~~~~~~~2051~~~~ \\
$[4]_{orb}[31]^{\sigma f}_{ts=21}~\bar{s}$        & ~~~~2103 ~~~~~~~~~2124 ~~~~~~~~~~~2090~~~~ \\
\hline
$ J^{\pi}=\frac{1}{2}^+$                         & I~~~~~~~~~~~~~~II~~~~~~~~~~~~~~III         \\
                                                 &       $(MeV)$               \\ \hline
$[31]_{orb}[4]^{\sigma f}_{ts=00}~\bar{s}$        & 2105 ~~~~~~~~~2016 ~~~~~~~~~~~1997         \\
$[31]_{orb}[4]^{\sigma f}_{ts=11}~\bar{s}$        & 2124 ~~~~~~~~~2051 ~~~~~~~~~~~2018         \\
~~~~~~~~~~~~~~~~~$(S=\frac{1}{2})$               &                          \\
$[31]_{orb}[4]^{\sigma f}_{ts=11}~\bar{s}$        & 2145 ~~~~~~~~~2070 ~~~~~~~~~~~2018         \\
~~~~~~~~~~~~~~~~~$(S=\frac{3}{2})$               &                          \\
$[31]_{orb}[4]^{\sigma f}_{ts=22}~\bar{s}$        & 2177 ~~~~~~~~~2122 ~~~~~~~~~~~2051           \\
\hline
\end{tabular}
\end{center}
\end{table}

\newpage

\section{ Conclusions.}

The structures of pentaquark states are studied by an adiabatic
approximation calculation in the chiral quark model. When the
interactions between $4q$ and $\bar{s}$ are considered, especially
the parameters in the annihilation interactions are fixed by
fitting the masses of $K$ and $K^*$,  our results show that state
$T=0, J^{\pi}=\frac{1}{2}^-$ is the lowest one, and its energy is
about $150-300 MeV$ higher than the $\Theta$'s mass. If we omit
the interactions between $4q$ and $\bar{s}$, then the state $T=0,
J^{\pi}=\frac{1}{2}^+$ can be lower then state $T=0,
J^{\pi}=\frac{1}{2}^-$, in agreement with what is claimed in
Ref.\cite{s6}. But the mass will be more than 400 MeV higher than
the observed $\Theta$ mass. This means that how to treat the
annihilation interaction reasonably is very important in the
calculation. On the other hand, even without omitting the
interactions between $4q$ and $\bar{s}$, all of the results of
various models with different parameters in this adiabatic
approximation calculation still give the lowest mass for the
$uudd$-$\bar s$ pentaquark at least more than 150 MeV above the
observed mass of $\Theta$. Furthermore, its $T=0,
J^{\pi}=\frac{1}{2}^-$ configuration has a potential s-wave KN
fall-apart mode \cite{s7} and hence is very difficult to explain
the very narrow width of the observed width of $\Theta$. It seems
that it is impossible to reproduce the observed low mass and
narrow width of $\Theta$ by quark models with reasonable model
parameters in the adiabatic approximation and a dynamical
calculation may be necessary for the further study.


\newpage



\end{document}